# Diels-Alder Reactions in Water are Determined by Microsolvation

Luis Ruiz Pestana[1,2,†], Hongxia Hao[1,2], and Teresa Head-Gordon[1,2*]

[1]*Chemical Sciences Division, Lawrence Berkeley National Laboratory*

[2]*Pitzer Center for Theoretical Chemistry, Departments of Chemistry, Bioengineering, and Chemical and Biomolecular Engineering*

*University of California, Berkeley, CA 94720*

**Abstract**

Nanoconfined aqueous environments and the recent advent of accelerated chemistry in microdroplets are increasingly being investigated for catalysis. The mechanisms underlying the enhanced reactivity in alternate solvent environments, and whether the enhanced reactivity due to nanoconfinement is a universal phenomenon, are not fully understood. Here, we use *ab initio* molecular dynamics simulations to characterize the free energy of a retro-Diels-Alder reaction in bulk water at very different densities and in water nanoconfined by parallel graphene sheets. We find that the broadly different global solvation environments accelerate the reactions to a similar degree with respect to the gas phase reaction, with activation free energies that do not differ by more than $k_bT$ from each other. The reason for the same acceleration factor in the extremely different solvation environments is that it is the microsolvation of the dienophile's carbonyl group that governs the transition state stabilization and mechanism, which is not significantly disrupted by either the lower density in bulk water or the strong nanoconfinement conditions used here. Our results also suggest that significant acceleration of Diels Alder reactions in microdroplets or on-water conditions can't arise from local microsolvation when water is present, but instead must come from highly altered reaction environments that drastically change the reaction mechanisms.

[†]*Department of Civil, Architectural, and Environmental Engineering, University of Miami, Coral Gables, FL 33146*

*Correspondence to: thg@berkeley.edu

**INTRODUCTION**

Inspired by the success of biological systems as catalysts,[1, 2] nanoconfinement has recently become a promising strategy to accelerate chemical reactions in engineered systems such as active sites of synthetic enzymes,[3] self-assembled nanocages,[4, 5] microdroplets,[6] or nanoporous materials.[7] Although confinement has been applied with relative success to some reactions, the general mechanisms underlying its catalytic effectiveness remain unclear, particularly for aqueous reactions, since nanoconfined water displays highly anomalous behavior with respect to the bulk.[8-10] For example, a study focused on prebiotic peptide chemistry in water nanoconfined by mineral surfaces found that not only the energetics of the reactions, but also the reaction mechanisms, can be affected by nanoconfinement.[11] Recently, Zare et al. and Cooks and co-workers have shown that many reactions can be accelerated in microdroplets, such as the spontaneous phosphorylation of sugars, production of ribonucleosides, and Diels-Alder chemistry.[6, 12-15] However, the underlying origin of the acceleration of the microdroplet remains unclear. Thus, predicting *a priori*, the effects of a particular alternative solvent environment on any specific chemical reaction, is a grand challenge scientific question.

The Diels-Alder (DA) reaction in bulk water is well understood and of great importance in organic chemistry. It was Breslow that first showed that the activation energy for cycloaddition of cyclopentadiene and methyl-vinyl-ketone drops by 3.8 kcal/mol in water with respect to the gas phase[16], while the retro-DA reaction has been found to be accelerated slightly further than the cycloaddition.[17] The experimentally observed acceleration was confirmed by Jorgensen and co-workers using Quantum Mechanics/Molecular Mechanics (QM/MM) approaches where the water molecules are described classically and the reactants quantum mechanically.[17, 18]

Although the computational studies consistently show that bulk water accelerates the reaction, at a quantitative level the results depend strongly on the QM level of theory employed. For example, for the reaction studied here, the activation barrier was shown to decrease by 2.8 kcal/mol when using the AM1 Hamiltonian,[17] but by 8.1 kcal/mol if the PDDG/PM3 semi-empirical molecular orbital theory was employed instead.[18] The major shortcoming of these early studies is that the water molecules were either treated implicitly[19, 20] or classically in the context of QM/MM[17, 18, 21], although more advanced treatments using a QM/QM' approach are a notable exception.[22, 23] Recent work has highlighted the necessity to fully capture the electronic degrees of freedom of the solvent to correctly describe the asynchronicity in electronic reorganization of the substrates in the transition state, which can be sensitive to the level of theory.[24]

In this work we study the retro-Diels-Alder (retro-DA) reaction between cyclopentadiene (CPD) and methyl-vinyl-ketone (MVK) in bulk water at both low and high density, as well under two different manifestations of nanoconfinement (Fig. 1). We simulate the entire system at the same level of QM theory, including the water molecules, with *ab initio* molecular dynamics (AIMD) based on density functional theory (DFT) using one of the best semi-local functional available, B97M-rV,[25, 26] which we have shown previously reproduces structural and dynamical properties of water at a level of accuracy on par with hybrid functionals.[27, 28]

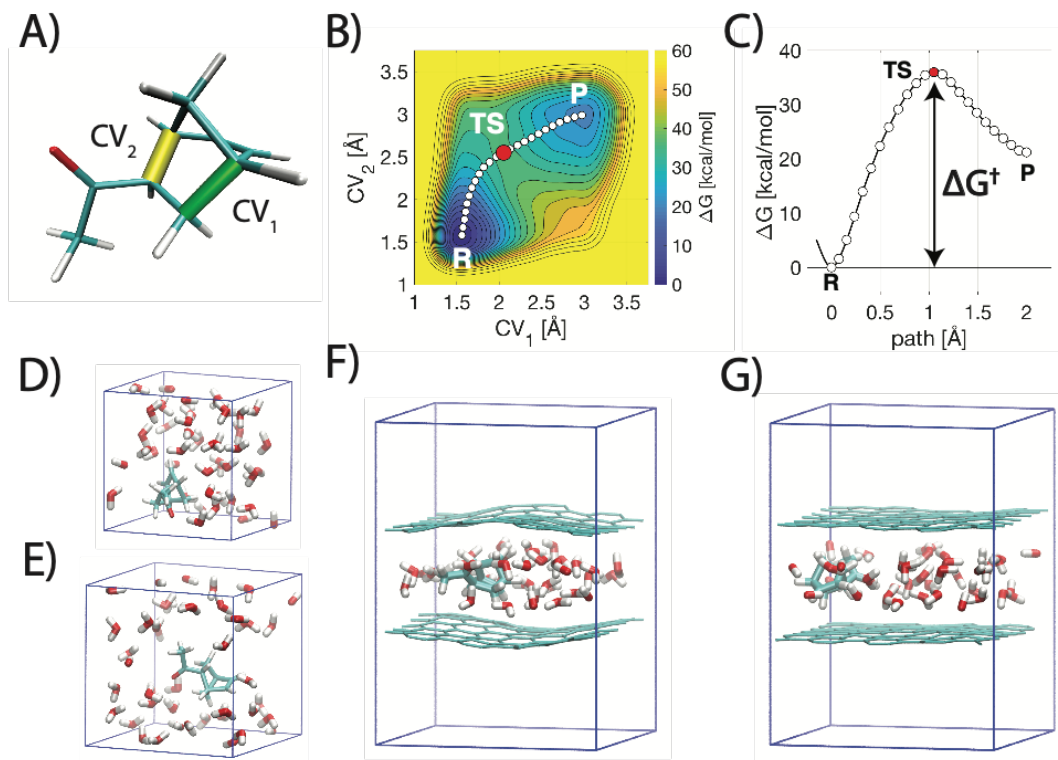

**Figure 1. The simulated CPD-MVK retro-Diels Alder reaction, computational approach, and the different environments.** (a) Cycloadduct between CPD and MVK. The collective variables $CV_1$ (green) and $CV_2$ (yellow) are the reactive bonds between diene and dienophile. (b) Representative free energy landscape, $\Delta G$, projected in the collective variable space. The white circles correspond to the minimum energy path (MEP) between the cycloadduct (reactant state) and the dissociated diene-dienophile products. The transition state (TS) is illustrated by a red circle. (c) Illustration of the relative free energy $\Delta G$ along the MEP and the activation free energy $\Delta G^\dagger$ of the reaction. The different simulated solvent environments: (d) *bulk-high* ($\rho = 1$ g/cm$^3$), (e) *bulk-low* ($\rho = 0.66$ g/cm$^3$), where $\rho$ is the mass density of the entire system, and (f) *conf-bent*, and (g) *conf-flat* with water nanoconfined between graphene sheets.

To characterize the CPD-MVK retro-DA reaction in the different solvent environments, we map the free energy surface (FES) along two collective variables in the gas phase, bulk water at two different densities, and in two different nanoconfined environments, using well-tempered metadynamics.[29] Similar to previous studies,[18, 21, 23, 30] we use as collective variables the two bonds between the CPD and MVK (Fig. 1a). After reconstructing the FES (Fig. 1b), we use the finite temperature string method (FTS) method to find the minimum energy path (MEP) between reactants and products (Fig. 1b-c). Finally, from the MEP, we calculate the activation free energy of the reaction in each environment $\Delta G^\dagger$ and the location of the transition state $(CV_1^\dagger, CV_2^\dagger)$. Further details regarding the different methodologies are provided in the Methods section; the specific dimensions of the simulation boxes and other details about the systems are provided in the Supporting Information.

The differences in the activation free energies for all the simulated cases are shown in Table 1. We find a decrease of the activation barrier in bulk water at normal mass density (i.e. *bulk-high*) with respect to the gas phase, $\Delta\Delta G^\dagger$, of approximately 4.6 kcal/mol, in line with previous experimental and computational studies and confirming the soundness of our methodology.[16-18, 21, 23, 30] Interestingly, the *bulk-low* condition displays similar acceleration rates despite the considerable differences in water density (0.66 g/cm$^3$ vs. 1 g/cm$^3$), disagreeing with previous results that the reaction rate would decrease at lower densities using QM/MM[18]. It is also clear that nanoconfinement does not change the Diels Alder reaction; for the reactions in nanoconfined water, both *conf-bent* and *conf-flat* systems, display similar $\Delta\Delta G^\dagger$, and are comparable to those observed in bulk water within $k_bT$. These results are remarkable in that extremely different solvation environments lead to nearly identical acceleration rates.

**Table 1. Free energies and reaction mechanism of the CPD-MVK under different environmental conditions.** Activation free energies ($\Delta G^\dagger$), relative activation free energies with respect to the gas phase ($\Delta\Delta G^\dagger$), location of the transition state in the collective variable space $(CV_1^\dagger, CV_2^\dagger)$ (Å), and asynchronicity ($\Delta CV^\dagger = CV_2^\dagger - CV_1^\dagger$) of the CPD-MVK retro-Diels-Alder reaction in the different environments. The averages and standard errors are calculated using a block average scheme on the aggregate data from the two independent runs in each environment. The convergence of $\Delta\Delta G^\dagger$ and $(CV_1^\dagger, CV_2^\dagger)$ is shown in Figures S1 and S2, respectively.

| System | $\Delta G^\dagger$ [kcal/mol] | $\Delta\Delta G^\dagger$ [kcal/mol] | $CV_1^\dagger$ [Å] | $CV_2^\dagger$ [Å] | $\Delta CV^\dagger$ [Å] |
|---|---|---|---|---|---|
| **gas phase** | 34.6 (0.1) | - | 2.02 (0.00) | 2.58 (0.00) | 0.56 (0.00) |
| **bulk-high** | 30.0 (0.1) | -4.6 (0.2) | 2.02 (0.00) | 2.73 (0.01) | 0.71 (0.01) |
| **bulk-low** | 29.8 (0.1) | -4.7 (0.2) | 2.06 (0.00) | 2.74 (0.01) | 0.69 (0.01) |
| **conf-bent** | 30.2 (0.3) | -4.3 (0.3) | 2.06 (0.01) | 2.73 (0.02) | 0.67 (0.02) |
| **conf-flat** | 30.5 (0.3) | -4.0 (0.3) | 2.03 (0.01) | 2.69 (0.01) | 0.65 (0.01) |

Furthermore, the mechanism for the CPD-MVK retro-DA reaction is well-characterized by the asynchronicity of the transition state, defined as the difference in length of the breaking bonds characterized by the two collective coordinates, $\Delta CV^\dagger = CV_1^\dagger - CV_2^\dagger$. We validated the location of the transition states for each case by reconstructing the isocommittor surface in collective variable space using a large number of unbiased simulations starting from multiple configurations close to the transition state. Details regarding this analysis are provided in the Methods section, but show excellent agreement between the calculated transition states and the 0.5 isocommittor line (Figure S3). Table 1 shows significant asynchronicity even in the gas phase, $\Delta CV^\dagger=0.56$, a value in excellent agreement with previous CBS-QB3 calculations[21] and results from higher-rung DFT functionals[24]. The results shown in Figure 2 and Table 1 also indicate that water, whether in bulk and regardless of density, or when nanoconfined, stabilizes the first bond-breaking event, which occurs along $CV_2$, while the second bond breaking event described by $CV_1$ remains approximately the same for all the systems. Although the nanoconfined environments exhibit slightly lower asynchronicity than the reactions in bulk water, the values are similarly elevated with respect to the gas phase, indicating no change in mechanism.

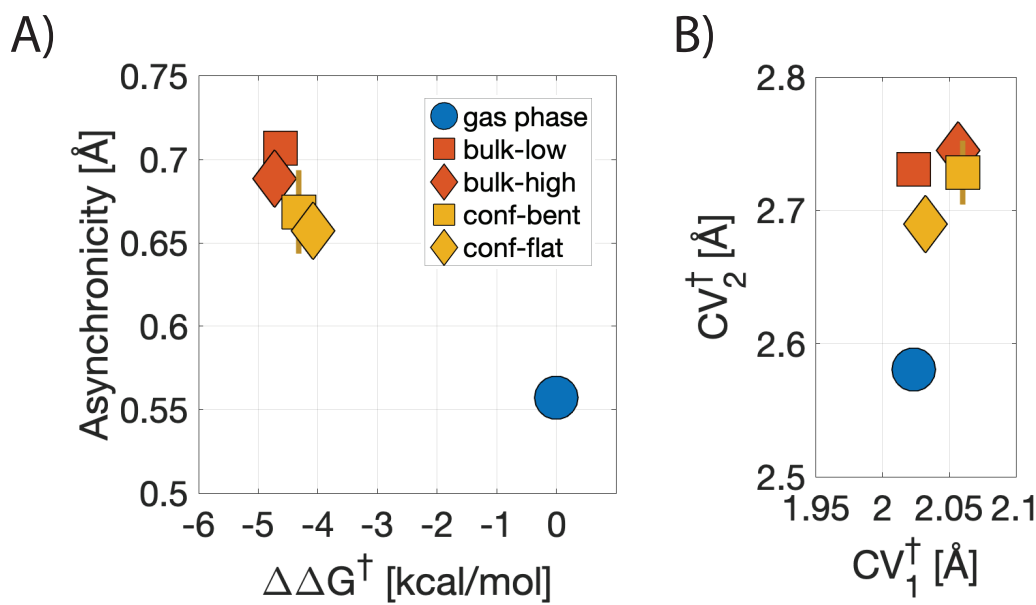

**Figure 2. Transition state asynchronicity for the CPD-MVK retro-DA reaction.** (a) Asynchronicity $\Delta CV^\dagger$ as a function of the activation free energy $\Delta\Delta G^\dagger$. (b) Values of the collective variables when the system is at the transition state. All environmental reaction conditions show similar activation free energies and mechanism within error bars.

To explain why these extremely different environments result in very similar activation free energies and the same reaction mechanism, we characterize the hydrogen bonding behavior of the

MVK's carbonyl group to water in the transition state with respect to both reactant and product states. Our results show that the number of hydrogen bonds is higher in the transition state than in the reactant or product states for all systems (Fig. 3a), which supports that the general catalytic asynchronous mechanism remains the same for all the cases.[17, 30] However, the overall number of water hydrogen bonds differs between solvation environments, with the bulk water system at the higher density forming the most hydrogen bonds, ~2.4 on average, which decreases to ~2.1 for the lower bulk density, while the nanoconfined environments hinder the hydrogen bonding to the carbonyl group further, down to ~1.9 and ~1.6 for the nanoconfined cases with bent and flat graphene sheets, respectively. These trends are supported by the diminishment in hydrogen-bond quality (Fig. 3b), in which the flat graphene system displays the longer and more bent hydrogen bonds. Finally, the difference in water hydrogen bond number and quality between reactant and transition states correlate well with the $\Delta\Delta G^\dagger$ (Fig. 3c and 3d), although the spread in the activation free energies is very small.

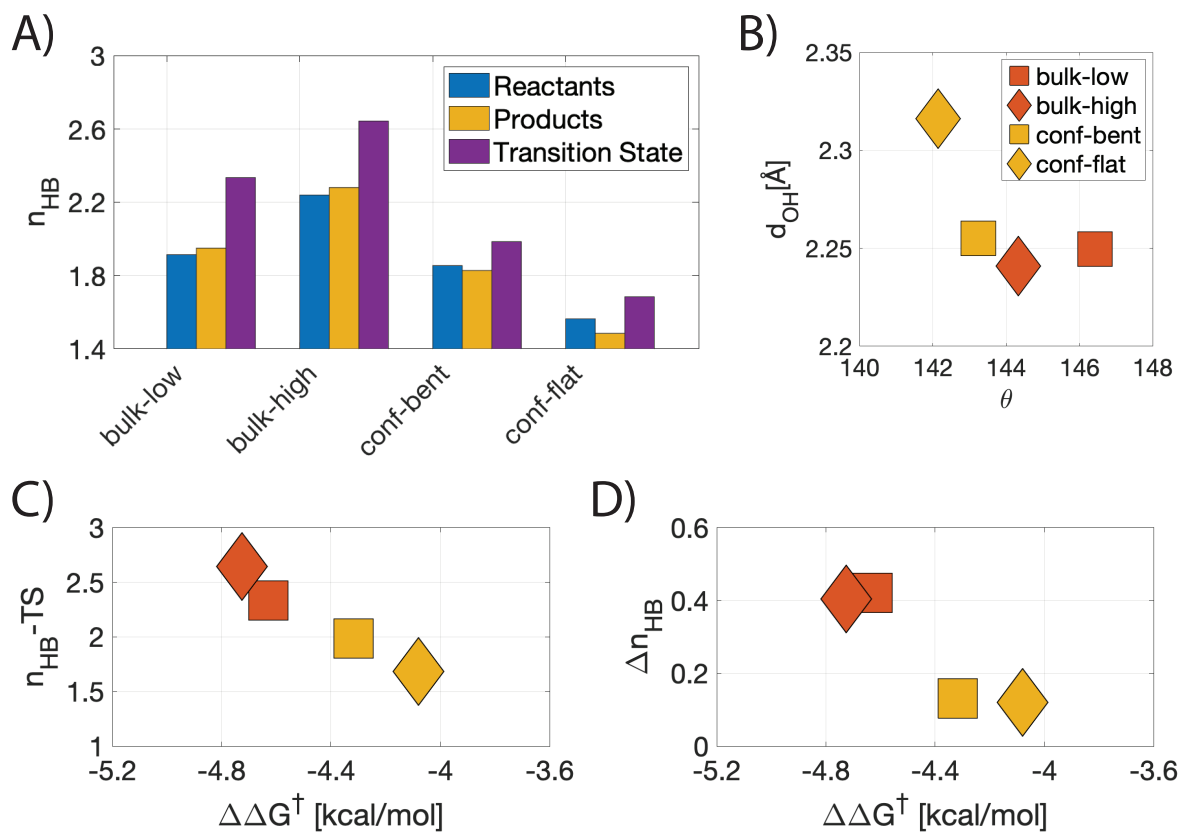

**Figure 3. Hydrogen bonding features of water near the carbonyl group of MVK.** (a) Number of hydrogen bonds when the systems are in the reactant, product, and transition states. (b) Average hydrogen bond distance, $d_{OH}$, vs. hydrogen bond angle, $\theta$; shorter and straighter hydrogen bonds (lower-right corner) are of higher quality. (c) Number of hydrogen bonds in the transition state as a function of $\Delta\Delta G^\dagger$. (d) Difference in hydrogen bonds between the reactant and transition states as a function of $\Delta\Delta G^\dagger$.

Our results highlight the remarkable robustness of Diels-Alder reactions in extreme environments. Even under very low bulk densities or under strong confinement conditions in which the average number and quality of water hydrogen bonds decreases, the reaction is accelerated at a relatively similar rate by the same mechanism. This highlights the role of the highly local "microsolvation" environment in catalysis, in this case only *one hydrogen bond, even of low quality,* is needed to accelerate the aqueous Diels-Alder reactions. Our findings indicate that almost any solvation environment with small amounts of water present cannot be used to accelerate the Diels Alder reaction further with respect to the standard bulk conditions at standard densities.

To place our results in the larger context of microdroplet chemistry, several factors have been proposed that can lead to greater reaction rates, including electrical voltage, salts, pH, electrochemical reactions, and concentration effects of reactants in rapidly evaporating drops. Reaction rate acceleration and/or mechanisms are also proposed to arise through interfacial or near-interfacial reactions[31-33], while others have suggested that the reactions could take place on the surface of electrosprayed water droplets or even in the gas-phase.[34-36] Furthermore, it's likely that the acceleration effects are system-specific and that several factors will influence the achievement of the large accelerations of several orders of magnitude observed in microdroplets. Decoupling these different effects remains tremendously challenging and is an active area of research.[35]

Relevant to this study, Zare and co-workers have found that only certain dienophile-diene reactants give rise to accelerations in ESI microdroplets that exceed bulk water reaction rates.[15] For other DA reactants they attribute the lack of acceleration or complete cessation of product formation to the insensitivity of the non-polar reactants to the microdroplet surface.[14] What we have shown is that for Diels Alder reactions, when they proceed through the microsolvation transition state mechanism, are accelerated at the same rate as the bulk water phase in almost any alternate environmental condition, and whether inside or at the surface of the aqueous microdroplet. This would imply that additional acceleration of the DA reaction when using electrospray microdroplets would indeed have to arise by concentration or new reaction mechanisms[11, 14]. Further studies are desired to ascertain why only certain DA reactions can be performed efficiently using "microdroplet chemistry" and what individual factors govern the accelerated rates observed.

In summary, this is the first computational study that focuses on Diels Alder reactions under nanoconfinement, a tremendously relevant topic in catalysis of organic reactions, and one of the reactions studied in the recent emergence of microdroplet chemistry[14, 15]. Our conclusions are unlike previous studies in which nanoconfinement is often thought to be a general strategy that can accelerate reactions[3-5, 7, 37], or previous conclusions that lower bulk water densities slow down the Diels Alder reaction[18]. We find that bulk water at both high and low densities, and strongly nanoconfined water, accelerate the Diels-Alder reaction to a remarkably similar degree with respect to the gas phase. We showed that the mechanism underlying the acceleration, which is the stabilization of the transition state through local water hydrogen bonding to promote the elongation of the first bond-breaking event, remains the same for all environments, including under strong nanoconfinement. For this reason, the implications for on-water and microdroplet chemistry for accelerated Diels Alder reactions[15, 33] must be a complete mechanistic change[11, 14], and possibly altering reactants so completely as to destroy the original reaction altogether[14].

## COMPUTATIONAL METHODS

***Simulated systems.*** We simulate a retro-Diels-Alder reaction between cyclopentadiene (CPD) and methyl-vinyl-ketone (MVK) (Fig. 1a). We simulate the reaction in five different environments: gas phase, bulk water at two different water densities ($\rho = 1$ g/cm$^3$ and $\rho = 0.66$ g/cm$^3$), and two different configurations of water nanoconfined by parallel graphene sheets; one where the graphene sheets remain flat, and another where the graphene sheets are slightly bent due to a small decrease in the size of the simulation box. We refer to the different environments as *gas phase*, *bulk-high* (Fig. 1d), *bulk-low* (Fig. 1e), *conf-bent* (Fig. 1f), and *conf-flat* (Fig. 1g) respectively.

***Ab initio Molecular Dynamics.*** Methodologically our work departs from previous Diels Alder studies that used QM/MM approximations by describing the entire system quantum mechanically at the same level of theory. We perform Born-Oppenheimer AIMD simulations based on DFT using the meta-GGA exchange-correlation functional B97M-rV, which captures non-local correlations necessary to describe dispersion interactions through the functional rVV10.[25, 26] We adopt the Gaussian plane-wave multigrid approach to DFT implemented in the subroutine QUICKSTEP[38] of the program CP2K.[39] To represent the core electrons, we use Goedecker-Teter-Hutter (GTH) pseudopotentials.[40, 41] In all the simulations, the self-consistent field cycle was converged using the orbital transformation method[42] to an accuracy of 5x10$^{-7}$. We use an energy cutoff of 400 Ry in all the simulations. Details regarding the size and composition of the simulated systems (Table S1), the basis sets employed (Table S2), and the total simulation times (Table S3) are given in the *Supporting Information*.

***Metadynamics.*** We perform both multiple-walker[43] and single-walker well-tempered[29] metadynamics-biased AIMD simulations. The simulations are all performed in the canonical NVT ensemble using a timestep of 0.5 fs. The temperature is set at 300 K and controlled using a velocity rescaling thermostat with a time constant of 200 fs. Before starting the metadynamics production simulations we equilibrate all the systems for at least 5 ps. After equilibration, we perform two independent production runs for each case. The specific duration of the different runs is shown in Table S3. The *temperature* parameter in the well-tempered metadynamics simulations is set to 5000 K, and the width of the deposited Gaussians is set to 0.07 Å. The maximum time interval between the deposition of two hills is 15 fs. For the single-walker simulations the height of the Gaussians is set to 6.27 kcal/mol. For the multi-walker simulations, we use 6 walkers and a height for the Gaussians of 0.627 kcal/mol. As collective variables we select the two bonds between CPD and MVK (Fig. 1a). In order to promote better the sampling of the areas of interest in the free energy landscape, harmonic walls with a stiffness of 40 kcal/mol act on both collective variables for values beyond 3 Å. We calculate the convergence and statistical error of both the activation energy barrier and the location of the transition state using a block average scheme applied to the time series of each quantity (Fig. S1, for the activation energy barrier and Figure S2 for the location of the transition state) We leave out of the analysis the first 40 ps of the time series and divide the rest in blocks of length $\tau_b$. We calculate the standard error $\sigma(\tau_b)$ of quantity $A$ as a function of the block size $\tau_b$ using:

$$\sigma(\tau_b) = \frac{s}{\sqrt{n_b}}$$

Where $s$ is the standard deviation:

$$s = \sqrt{\frac{1}{n_b - 1}\sum_{i=1}^{n_b}(\langle A\rangle_b - \langle A\rangle_{sim})^2}$$

and $\langle A\rangle_b$ is the average of quantity $A$ for each block of length $\tau_b$:

$$\langle A\rangle_b = \frac{1}{\tau_b}\sum_{i=1}^{\tau_b} A_i$$

*Finite Temperature String Method.* We use the finite temperature string (FTS) method[33] to find the minimum energy path (MEP) in the free energy surface (FES) reconstructed from the metadynamics-biased AIMD simulations. We use the forward Euler method to evolve the over-damped dynamics of the string images in collective variable space:

$$(CV_1)_i = (CV_1)_{i+1} - h\nabla_{CV_1}G + \eta\sqrt{2h\gamma}$$
$$(CV_2)_i = (CV_2)_{i+1} - h\nabla_{CV_2}G + \eta\sqrt{2h\gamma}$$

Where $G$ is the free energy surface, $\eta$ is a white-noise driver, the timestep $h = 5 \times 10^{-3}$, and the effective temperature $\gamma = 0.01$. We use 36 images along the path joining reactants and products (see white circles in Fig. 1b). Initially, we relax the string for 5,000 steps. We then collect data over the next 5,000 steps to determine the MEP. Details regarding the FTS method can be found references [44] and [45].

*Committor Analysis.* To validate location of the transition state and to get information about the shape of the committor surface in the region around it, we performed a large number of unbiased simulations long enough to reach either the reactant (i.e. the cycloadduct) or the product (i.e. the dissociated diene-dienophile). For each system, we choose at least 24 different initial configurations of the system around the transition state. From each of these initial configurations, we launch 24 short unbiased simulations initialized with random Boltzmann distributed velocities but otherwise identical, and assigned a committor value between 0 and 1 depending on the fraction of trajectories that end up in the product state. We reconstruct the committor surface by interpolating the committor values at those initial configurations. Finally, we calculate the 0.5 isocommittor line on those surfaces. The results of the committor analysis displayed in Figure S3 show a very good agreement. The unbiased trajectories in CV space and the points that were chosen as initial configurations for each chase are also shown in Figure S3.

**ACKNOWLEDGEMENTS**. We thank the U.S. DOE under the Basic Energy Sciences CPIMS program, Contract No. DE-AC02-05CH11231 for research support. This research used computational resources of the National Energy Research Scientific Computing Center, a DOE Office of Science User Facility supported by the Office of Science of the U.S. Department of Energy under Contract No. DE-AC02-05CH11231, under an ASCR Leadership Computing Challenge (ALCC) award.

**SUPPORTING INFORMATION**. Description of methods used to perform calculations, statistical summary of all data, and underlying trajectory information. The Supporting Information is available free of charge on the ACS Publications website.

**FOR TOC ONLY**

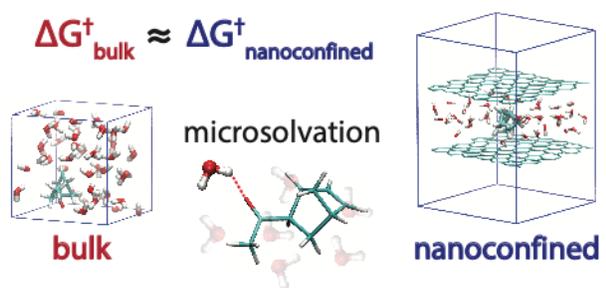